\documentclass[aps,twocolumn,superscriptaddress,nofootinbib]{revtex4-2}

\usepackage[dvipsnames]{xcolor}
\usepackage{hyperref}
\hypersetup{
  breaklinks=true,
  colorlinks=true,
  allcolors=BlueViolet,
}

\usepackage{graphicx} 

\usepackage{physics,braket,bm}
\renewcommand{\set}[1]{\left\{#1\right\}}
\newcommand{\ii}{\mathrm{i}\mkern1mu} 

\newcommand{\CTP}{\texttt{CTP}}
\newcommand{\CTPO}{\texttt{CTP}$_{\mathrm{old}}$}
\newcommand{\CTPF}{\texttt{CTP}$_{\mathrm{fixed}}$}
\newcommand{\CTPN}{\texttt{CTP}$_{\mathrm{new}}$}
\newcommand{\SE}{\texttt{SE}}
\newcommand{\EC}{\texttt{EC}}

\newcommand{\PP}[1]{\texttt{P}$_2^{#1}$}
\newcommand{\DNM}[1]{\texttt{DNM}$^{#1}$}

\renewcommand{\d}{\texttt{-}}
\newcommand{\CO}{\texttt{C}$_{\mathrm{old}}$}
\newcommand{\CN}{\texttt{C}$_{\mathrm{new}}$}

\newcommand{\CnNOT}[1]{\texttt{C$_{#1}$NOT}}
\newcommand{\CNOTn}[1]{\texttt{CNOT$_{#1}$}}
\newcommand{\CnZ}[1]{\texttt{C$_{#1}$Z}}
\newcommand{\CZn}[1]{\texttt{CZ$_{#1}$}}

\newcommand{\CkNOT}{\CnNOT{k}}
\newcommand{\CNOTk}{\CNOTn{k}}
\newcommand{\CkZ}{\CnZ{k}}
\newcommand{\CZk}{\CZn{k}}

\newcommand{\RZ}{\texttt{RZ}}

\let\log\relax
\newcommand{\log}{\mathrm{log}}
\newcommand{\phys}{\mathrm{phys}}
\newcommand{\cor}{\mathrm{cor}}
\newcommand{\fault}{\mathrm{fault}}

\newcommand{\F}{\mathcal{F}}

\newcommand{\EE}{\mathbb{E}}

\newcommand{\Om}[1]{\Omega_{\mathrm{#1}}}
\DeclareMathOperator{\round}{round}
\DeclareMathOperator{\argmax}{argmax}

\usepackage{amssymb}

\usepackage[inline]{enumitem}
\setlist[enumerate]{leftmargin=*}

\newcommand{\InfleqtionC}{Infleqtion, Inc., Chicago, IL, 60615, USA}
\newcommand{\InfleqtionM}{Infleqtion, Inc., Madison, WI, 53703, USA}
\newcommand{\UWM}{Department of Physics, University of Wisconsin-Madison, 1150 University Avenue, Madison, WI, USA}
\newcommand{\Kavli}{Kavli Institute for Theoretical Sciences, University of Chinese Academy of Sciences, Beijing 100190, China}

\begin{document}

\title{Fault-tolerant measurement-free quantum error correction with multi-qubit gates}
\author{Michael A.~Perlin}
\email{mika.perlin@gmail.com}
\affiliation{\InfleqtionC}
\author{Vickram N.~Premakumar}
\affiliation{\UWM}
\author{Jiakai Wang}
\affiliation{\UWM}
\author{Mark Saffman}
\affiliation{\UWM}
\affiliation{\InfleqtionM}
\author{Robert Joynt}
\affiliation{\UWM}
\affiliation{\Kavli}
\date{\today}

\begin{abstract}
  Measurement-free quantum error correction (MFQEC) offers an alternative to standard measurement-based QEC in platforms with an unconditional qubit reset gate.
  We revisit the question of fault tolerance (FT) for a measurement-free variant of the Steane code that leverages multi-qubit gates and redundant syndrome extraction, finding previously overlooked phase-flip errors that undermine FT.
  We then construct a revised MFQEC circuit that is resistant to all single-qubit errors, but which nonetheless cannot tolerate certain correlated errors.
  In order to investigate FT systematically, we introduce an efficient method to classically simulate MFQEC circuits with (i) Clifford gates for syndrome extraction, (ii) syndrome-controlled Pauli operations for decoding, and (iii) a Pauli noise model.
  We thereby find a pseudothreshold of $\sim0.7\%$ for our revised MFQEC Steane code under a restricted noise model previously considered in the literature.
  We then relax noise model assumptions to identify general requirements for FT with multi-qubit gates, finding that existing multi-qubit neutral atom gates are incompatible with fault-tolerant syndrome extraction in a straightforward implementation of both measurement-based and measurement-free variants of the Steane code.
  Decomposing multi-qubit gates into two-qubit gates similarly spoils FT.
  Finally, we discuss the theoretical ingredients that are necessary to recover FT for MFQEC codes, including single-shot FT and a recent proposal by Heu{\ss}en \textit{et al.}~[arXiv:2307.13296] to achieve FT by ``copying'' errors onto an ancilla register.
  By combining multi-qubit gates, redundant syndrome extraction, and copy-assisted FT, we construct a measurement-free and fault-tolerant variant of the Steane code with a pseudothreshold of $\sim0.1\%$.
\end{abstract}

\maketitle

\section{Introduction}

Large-scale quantum computers will require quantum error correction (QEC) for reliable computation.
The choice of best QEC protocol will depend strongly on the physical platform of a given quantum computer.
The dominant paradigm for QEC relies heavily on repeated measurements to mitigate the buildup of entropy (errors) throughout a quantum computation.
However, this reliance on measurements introduces its own set of challenges.
In quantum computers based on neutral atoms, for example, measurement time is much longer than other gate times, and crosstalk due to light scattering over the course of a measurement can be difficult to control \cite{beterov2015rydberg}.
For semiconductor quantum dot qubits (e.g.,~spin qubits), long measurement times can similarly make traditional QEC less appealing \cite{petta2005coherent, thorgrimsson2017extending}.
At the same time, these platforms allow for other non-unitary operations such as an unconditional qubit reset, which theoretically enables QEC.
Measurement-free quantum error correction (MFQEC) has therefore been proposed as an alternative to the standard measurement-based paradigm \cite{paz-silva2010fault, li2012recovery, crow2016improved, ercan2018measurementfree, inada2021measurementfree, heussen2023measurementfree}.

In a nutshell, standard qubit-based QEC can be broken down into error correction cycles that consist of four steps:
\begin{enumerate*}[label=(\roman*)]
  \item syndrome extraction, wherein the values of certain data-qubit observables that can diagnose errors are ``written'' onto the state of ancilla qubits,
  \item \label{step:measure_QEC}
    syndrome (ancilla qubit) measurement,
  \item classical decoding of the measurement results to infer any errors that may have occurred, and
  \item \label{step:correct_QEC}
    error correction on the data qubits%
    \footnote{
      Depending on the QEC code being used, the error correction step may forego performing physical operations on the quantum computer, and instead update a virtual ``frame'' that defines the computational basis for the data qubits.
    }.
\end{enumerate*}
MFQEC essentially replaces steps \ref{step:measure_QEC}--\ref{step:correct_QEC} by a \emph{coherent decoding} step that pumps entropy out of the data qubits into the ancilla qubits, followed by an unconditional dissipative ``reset'' of the ancilla qubits to fiducial $\ket{0}$ states, without having to perform any measurements.
The coherent decoding step of MFQEC is the reason why it is also sometimes referred to as coherent error correction.
An alternative QEC paradigm is provided by \emph{autonomous} or \emph{dissipative} QEC, which forgoes the need for measurement through continuous engineered dissipation into a logical code space \cite{ahn2002continuous, sarovar2005continuous, sarovar2005continuous, reiter2017dissipative, li2018unconventional, lebreuilly2021autonomous}.
We focus on gate-based (discrete) MFQEC in this work.

A QEC protocol, whether measurement-based or measurement-free, is said to be \emph{fault-tolerant} if, when the QEC protocol itself consists of error-prone operations, the logical qubit encoded by the QEC code can achieve better performance in a quantum algorithm than the underlying physical qubits.
By conventional wisdom, the overheads required for fault-tolerant MFQEC are thought to be significantly greater than those for fault-tolerant measurement-based QEC \cite{paz-silva2017multiqubit}.
However, some recent works have challenged this conventional wisdom \cite{crow2016improved, ercan2018measurementfree, inada2021measurementfree}, which may make MFQEC more attractive for some quantum computing platforms.
Notably, Ref.~\cite{crow2016improved} found MFQEC pseudothresholds, or physical error rates required to achieve parity between physical and logical qubit performance, that are comparable to pseudothresholds in measurement-based codes.
A key enabling factor for favorable MFQEC pseudothresholds is the availability of native multi-qubit gates, which circumvents the need to perform expensive decompositions of multi-qubit decoding logic into a two-qubit gate set.
In principle, all necessary multi-qubit gates have been previously proposed in neutral atom platforms \cite{wu2010implementation, isenhower2011multibit, levine2019parallel, young2021asymmetric, pelegri2022highfidelity, farouk2023parallel, kinos2023optical}.

However, Ref.~\cite{crow2016improved} overlooked a class of physically relevant single-qubit errors that have significant consequences for code performance.
Specifically, the simulations in Ref.~\cite{crow2016improved} did not allow for phase-flip errors on ancilla qubits during syndrome extraction.
Including these errors spoils FT for the MFQEC circuits that were considered.
In other words, the previously considered circuits have no pseudothreshold: their logical qubit error rates are lower bounded by physical error rates.

Focusing on the Steane code as an exemplar, in this work we show that an amendment to the circuit design in Ref.~\cite{crow2016improved} recovers FT with respect to all single-qubit errors.
We then develop a classical algorithm to efficiently simulate a broad class of MFQEC circuits with a Pauli noise model, using which we validate the new circuit design and demonstrate the existence of a pseudothreshold.
In addition, we argue that certain features of the previously considered noise model are in fact too \emph{pessimistic}, such that we altogether find a more favorable pseudothreshold of $\sim0.7\%$ for the MFQEC Steane code with the noise model considered in Ref.~\cite{crow2016improved}.

However, the noise model that admits a pseudothreshold for the MFQEC Steane code makes certain assumptions about correlated data qubit errors.
We therefore relax these assumptions to perform a systematic investigation of the requirements for FT.
We thereby identify a class of correlated errors that spoil FT for the new circuit design, and observe that these errors are relevant in all protocols (that we can find) to natively implement the corresponding multi-qubit gates with neutral atoms.
Specifically, we note the absence of single control, multi-target \CNOTk{}-like gate protocols that are compatible with fault-tolerant syndrome extraction in a straightforward implementation of both measurement-based and measurement-free variants of the Steane code, and discuss how the underlying problem might be avoided with higher-distance QEC codes.
We then consider the possibility of decomposing multi-qubit gates down to a two-qubit gate set, and find that doing so similarly spoils FT.
Finally, we discuss general requirements for fault-tolerant MFQEC with both multi-qubit and two-qubit gate sets, highlighting two avenues in particular: single-shot FT, and a recent proposal to achieve FT by ``copying'' errors onto an ancilla register \cite{heussen2023measurementfree}.
The recent proposal in particular allows us to construct a fault-tolerant variant of the Steane code with a pseudothreshold of $\sim0.1\%$.

The remainder of this paper is structured as follows.
In Section \ref{sec:design}, we revisit the MFQEC circuit design in Ref.~\cite{crow2016improved}, discuss errors that spoil its FT, and present a revised circuit design that recovers FT.
We present an efficient classical algorithm for simulating a class of MFQEC codes with Pauli noise models in Section \ref{sec:simulation}.
We then analyze the FT properties of our revised circuit design in Section \ref{sec:fault_tolerance}, and conclude in Section \ref{sec:summary}.

\section{Circuit design}
\label{sec:design}

\begin{figure*}
  \centering
  \includegraphics[width=\linewidth]{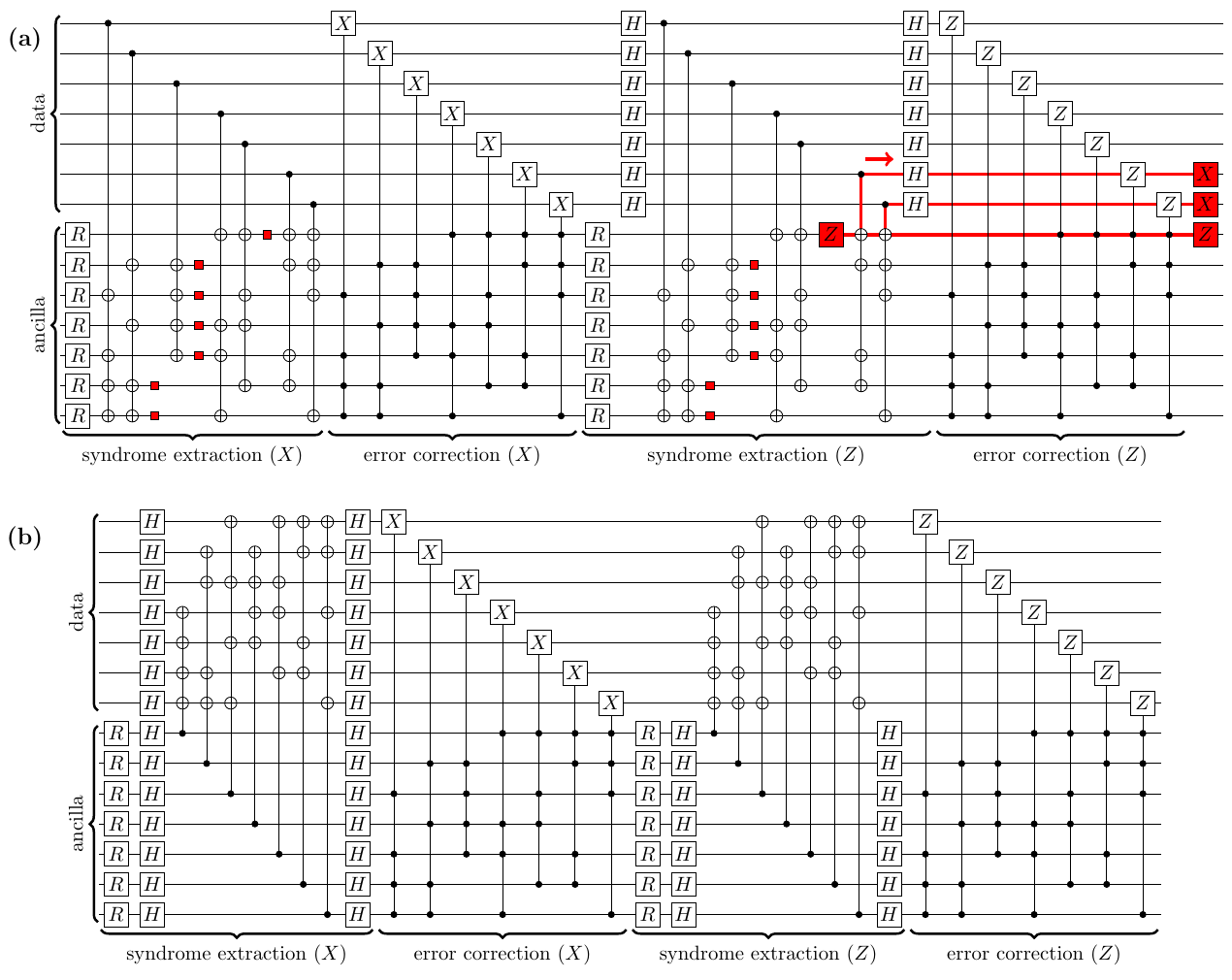}
  \caption{
    \textbf{(a)}
    The MFQEC circuit with redundant syndrome extraction proposed for the Steane code in Ref.~\cite{crow2016improved}.
    Here $H$ is a Hadamard gate and $R$ is a non-unitary \texttt{Reset} gate that unconditionally forces a qubit into the $\ket{0}$ state.
    The top half of the circuit addresses data qubits, and the bottom half addresses ancilla qubits.
    This circuit is not fault-tolerant: red squares ({\tiny\color{red}{$\blacksquare$}}) indicate locations at which a single-qubit phase error on an ancilla qubit propagates to an uncorrectable error on the data qubits.
    As a concrete example, we trace (with thicker red lines) the propagation of a single-qubit phase ($Z$) error on the top ancilla qubit to a two-qubit $XX$ error on the data qubits.
    \textbf{(b)}
    Revised MFQEC circuit for the Steane code, which can now tolerate arbitrary single-qubit errors.
    The primary change from circuit (a) is reversing the roles that data and ancilla qubits play at the syndrome extraction steps.
    This reversal limits the number of times that each ancilla interacts with data qubits, and thereby prevents single-qubit ancilla errors from propagating to uncorrectable data errors.
    The simulations in Section \ref{sec:fault_tolerance} were performed with \CZn{4} gates at the $X$-type syndrome extraction stage, rather than Hadamard-transformed \CNOTn{4} gates.
  }
  \label{fig:circuits_DBR}
\end{figure*}

Figure \ref{fig:circuits_DBR}(a) shows a circuit introduced in Ref.~\cite{crow2016improved} to implement a single round of error correction for a MFQEC variant of the Steane code \cite{steane1997multipleparticle}.
Data and ancilla qubits are, respectively, addressed by the top and bottom halves of this circuit.
This circuit can be divided into four stages, each of which implements syndrome extraction or coherent error correction for bit-flip ($X$-type) or phase-flip ($Z$-type) errors on the data qubits.

\begin{figure}
  \centering
  \includegraphics{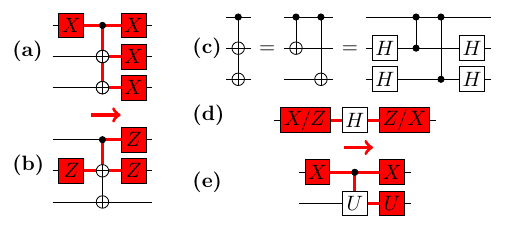}
  \caption{
    Propagation of \textbf{(a)} bit-flip ($X$) and \textbf{(b)} phase-flip ($Z$) errors through \CNOTk{} gates.
    Note that \CNOTk{} gates commute with phase-flip errors on their control qubits and bit-flip errors on their target qubits.
    \textbf{(c,d,e)} The elementary identities that can be used to derive \textbf{(a,b)}.
  }
  \label{fig:prop_basic}
\end{figure}

An important feature of the circuit in Figure \ref{fig:circuits_DBR}(a), in contrast to the measurement-based Steane code, is that of \emph{redundant} syndrome extraction, whereby an overcomplete set of stabilizers (error-witnessing observables) are written onto ancilla qubits.
Redundant syndrome extraction is generally necessary for MFQEC circuits to tolerate faulty syndrome readout, or bit-flip errors on the ancilla qubits%
\footnote{
  In measurement-based QEC, syndrome readout errors are tolerated through repeated rounds of syndrome extraction and measurement prior to error correction.
}.
The choice of redundancy scheme can be optimized using design theory \cite{premakumar20212designs}.
However, the circuit in Figure \ref{fig:circuits_DBR}(a) cannot tolerate certain phase-flip errors on the ancilla qubits during syndrome extraction.
Specifically, back-action causes some ancilla phase-flip errors to propagate through \CNOTk{} operations to two-qubit $XX$ or $ZZ$ errors on the data qubits at the end of a code cycle.
These data qubit errors are uncorrectable by the Steane code, and therefore irreversibly corrupt the state of an encoded logical qubit.
The probability of logical state corruption in one code cycle is thereby bounded from below by the probability of an intolerable phase-flip error on the ancilla qubits.
For reference, we provide the identities used to propagate errors through \CNOTk{} gates in Figure \ref{fig:prop_basic}.

Having identified the intolerable errors for the circuit in Figure \ref{fig:circuits_DBR}(a), we can reconfigure the circuit to avoid these faults, thereby arriving at the new circuit in Figure \ref{fig:circuits_DBR}(b).
The new circuit uses a design-based redundancy scheme for syndrome extraction \cite{premakumar20212designs}, which is maximally tolerant of ancilla qubit errors and does not allow any single-qubit error to propagate to an uncorrectable error on the the data qubits.
The crucial operational change from the circuit in Figure \ref{fig:circuits_DBR}(a) is to switch the roles of the data and ancilla qubits in the syndrome extraction steps.
Specifically, the new circuit can be obtained from the old by splitting data-controlled \CNOTk{} gates into $k$ \texttt{CNOT} gates, reversing the role of control and target qubits with Hadamard gates, and recombining the transformed \texttt{CNOT} gates into ancilla-controlled \CNOTk{} gates.
Each ancilla qubit now interacts with data qubits only once per syndrome extraction step, as the control qubit for a controlled-stabilizer operation.
This pattern of interaction prevents ancilla qubit errors from propagating to uncorrectable data qubit errors.
Of course, switching the roles of the data and ancilla qubits means that data qubit errors can now propagate to high-weight errors on the ancilla qubits.
However, design-based redundancy allows the code to tolerate these errors.

\section{Circuit simulation}
\label{sec:simulation}

The circuits in Figure \ref{fig:circuits_DBR} are difficult to simulate directly.
Here, we present an efficient method to classically simulate the effect of a MFQEC circuit with mid-circuit Pauli errors.
This method can, in turn, be used as subroutine in a randomized sampling algorithm to compute the \emph{logical error rate} of a noisy MFQEC circuit, which we define as the probability that the circuit irreversibly corrupts a logical state of its data qubits.
We describe the details of a particular algorithm based on importance sampling, similar to that in Refs.~\cite{li2017fault, trout2018simulating}, in Appendix \ref{sec:sampling}, which is used for the analysis in Section \ref{sec:fault_tolerance}.

It is worth briefly considering why a straightforward numerical analysis of the circuits in Figure \ref{fig:circuits_DBR} is challenging in the first place.
After all, these circuits address 14 qubits, whose pure state has a memory footprint of only $2^{14}\approx 10^4$ complex numbers.
The introduction of noise, however, means that we must consider \emph{mixed} states of 14 qubits, giving rise to density matrices with memory footprint $(2^{14})^2 = 2^{28} \approx 10^9$.
Finally, features of the code such as a logical error rate are in fact features of a \emph{quantum channel} that maps density operators to density operators, and therefore has a memory footprint of $(2^{28})^2 = 2^{56} \approx 10^{17}$ complex numbers, or $\sim1$ exabyte with standard (64-bit) floating-point precision.
This large memory footprint makes a direct numerical calculation of logical error rates infeasible.
Ref.~\cite{crow2016improved} tried to circumvent this difficulty by treating ancilla qubits classically, but this treatment neglects ancilla-qubit phase errors that spoil FT.

Standard strategies for numerically analyzing QEC circuits leverage the Gottesman-Knill theorem \cite{gottesman1998heisenberg, aaronson2004improved} for efficient classical simulation of Clifford circuits, which enables state-of-the-art tools such as \texttt{Stim} \cite{gidney2021stim} to simulate circuits with thousands of qubits and millions of operations in a matter of seconds.
However, the \CkNOT{} operations in Figure \ref{fig:circuits_DBR} are non-Clifford, which precludes the use of Clifford circuit simulators.
Nonetheless, it is still possible to classically simulate MFQEC circuits such as those in Figure \ref{fig:circuits_DBR} with a Pauli noise model.
The only requirements are that the MFQEC circuit uses Clifford gates for syndrome extraction onto ancilla qubits, and syndrome-controlled Pauli operations for decoding and error correction%
\footnote{
  The simulation method presented here is straightforward to generalize to the case of syndrome-controlled Clifford operations, but doing so is unnecessary for this work.
}.

\begin{figure}
  \centering
  \includegraphics{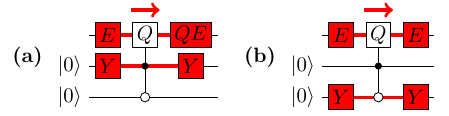}
  \caption{
    Examples of a Pauli error \textbf{(a)} is and \textbf{(b)} is not mutated when propagating forward through a multi-controlled gate that conditionally applies the Pauli string $Q$.
    Here $E$ is an arbitrary Pauli string.
    For the class of circuits we consider, the Pauli error being propagated forward is always preceded by a noiseless syndrome extraction subcircuit that leaves the control qubits in $\ket{0}$, allowing the conditions of a multi-controlled operation to be evaluated with classical logic.
  }
  \label{fig:prop_mcg}
\end{figure}

Two key observations enable the efficient simulation of noisy MFQEC circuits.
First, that \textit{noiseless} syndrome extraction subcircuits are Clifford, and have a known, trivial effect: they leave the logical state of the data qubits unaffected and prepare syndrome qubits in $\ket{0}$.
Second, that the non-Clifford gates in a MFQEC circuit are conditional Pauli strings controlled by the state of ancilla qubits that store a syndrome.
The general simulation strategy is therefore to sweep through a circuit to propagate Pauli errors forward, and classically evaluate syndrome-controlled logic along the way.
Altogether, there are three cases to consider for propagating Pauli errors forward through a circuit operation:
\begin{enumerate}
  \item \textbf{Unitary Clifford gates.}
    By definition, a Pauli string $P$ can be propagated forward through a unitary Clifford gate $g$ by mutating $P$ into another Pauli string $P_g = g P g^\dag$, since $P_g g = g P g^\dag g = g P$.
  \item \textbf{Reset gates.}
    All single-qubit Pauli errors annihilate against \texttt{Reset} gates.
  \item \textbf{Multi-controlled Pauli strings.}
    By assumption, every multi-controlled operation $M$ conditionally applies a Pauli string $Q_M$, depending on the state of control qubits that store a syndrome.
    A noiseless syndrome extraction circuit preceding $M$ leaves these control qubits in the classical state $\ket{0}$.
    If a Pauli error $P$ immediately precedes $M$ (and if $P$ is not preceded by any other errors) then we can then classically determine whether $M$ will apply $Q_M$ and mutate $P\to Q_M P$ or $P\to P$ accordingly (see Figure \ref{fig:prop_mcg}).
    Specifically, $M$ applies $Q_M$ if and only if $P$ addresses every closed control of $M$ with a bit-flip ($X$ or $Y$) error, and $P$ \textit{does not} address \textit{any} open control of $M$ with a bit-flip error%
    \footnote{
      Here ``closed'' and ``open'' controls respectively mean that an operation is conditioned on the corresponding control qubit being in the state $\ket{1}$ and $\ket{0}$.
    }.
\end{enumerate}
If a circuit initially contains multiple Pauli errors, the first (earliest) Pauli error should be propagated forward and merged with subsequent errors as they are encountered.
Altogether, since a noiseless QEC code cycle has no net effect on its data qubits, the final Pauli error propagated to the end of a circuit is equal to the net effect of the original circuit (with fixed mid-circuit errors) on its qubits.
We note that this algorithm relies on the correctness of syndrome extraction subcircuits, which must be verified independently.

\section{Fault tolerance analysis}
\label{sec:fault_tolerance}

The new circuit in Figure \ref{fig:circuits_DBR}(b) for a measurement-free error correction cycle of the Steane code restores FT with respect to arbitrary single-qubit errors%
\footnote{
  Here by ``fault tolerance'', we simply mean the existence of a pseudothreshold at which the logical error rate is equal to a ``physical'' error rate associated with the underlying qubits.
}.
However, a restriction to single-qubit errors is unrealistic in the presence of multi-qubit gates that may induce multi-qubit errors.
Here, we analyze the FT properties of this circuit more closely and identify the conditions that a noise model must satisfy to preserve FT.
Notably, we find that existing proposals for multi-qubit \CkNOT{} gates with neutral atoms do not satisfy the requirements for FT.
Finally, we consider the possibility of decomposing multi-qubit gates down to a two-qubit gate set, and discuss the prospects for fault-tolerant MFQEC with a two-qubit gate set.

\subsection{Noise models}

We will primarily consider five noise models.
Each of these models corresponds to the insertion of (possibly multiple) Pauli channels immediately after each gate in a circuit.
Every Pauli channel has the same probability $p_\phys$ of applying an error on its qubits.
For simplicity, we neglect memory (idle) errors and state preparation errors after reset gates, which is equivalent to modifying single-qubit error rates in the noise models below, and has no substantial consequence for our results.
Memory errors are also negligible compared to gate errors in, e.g., neutral atom platforms.

The five noise models that we consider are:
\begin{enumerate}
  \item \CTPO{}: The same noise model for gates as in Ref.~\cite{crow2016improved} (described below).
    We stress that this model is not realistic, as it does not include phase errors on ancilla qubits.
  \item \CTPF{}: A ``fixed'' version of \CTPO{}, which additionally allows for phase errors on ancilla qubits.
  \item \CTPN{}: A further modified version of \CTPF{}, for which the ``physical error rate'' $p_\phys$ is the probability of an error after each gate.
  \item \PP{}: A noise model that allows for all single- and two-qubit Pauli errors.
  \item \DNM{}: A fully depolarizing noise model.
\end{enumerate}

The letters ``CTP'' in \CTPO{}, \CTPF{}, and \CTPN{} are shorthand for ``control-target pairs'', reflecting the types of errors that these noise models allow after multi-qubit gates.
The \CTPF{} noise model inserts a single-qubit depolarizing channel after a single-qubit gate, and a two-qubit depolarizing channel on every control-target qubit pair after a multi-qubit gate.
That is, \CTPF{} inserts a depolarizing channel on qubits $q_1$ and $q_2$ after the multi-qubit gate $G$ iff one of $q_1$ and $q_2$ is a control qubit and the other a target qubit in $G$.
The gates \CnNOT{4} and \CNOTn{4}, for example, would both be followed by four two-qubit depolarizing channels.
The \CTPO{} noise model is identical to \CTPF{}, but replaces depolarizing channels with Pauli channels that exclude phase errors (i.e., a Pauli-$Z$ or Pauli-$Y$ error) on ancilla qubits while keeping the overall probability of error fixed.

The \CTPN{}, \PP{}, and \DNM{} noise models all insert only a single Pauli channel after each gate.
In the event of an error for this channel, occurring with probability $p_\phys$, the channel uniformly picks a Pauli string to apply to its qubits.
The difference between \CTPN{}, \PP{}, and \DNM{} is the set of errors that are ``allowed''.
\CTPN{} allows for the same errors as \CTPO{} and \CTPF{}: all single-qubit Pauli errors, and all two-qubit Pauli errors on control-target qubit pairs.
\PP{} allows for all single- and two-qubit Pauli strings, and \DNM{} allows for all Pauli strings.
In all of these cases, the overall probability of no error after a gate, which for a Pauli noise model is equal to the entanglement fidelity \cite{nielsen1996entanglement, nielsen2002simple} of that gate, is $1-p_\phys$.

\subsection{Recovering fault tolerance}
\label{sec:CTP_sims}

\begin{figure}
  \centering
  \includegraphics{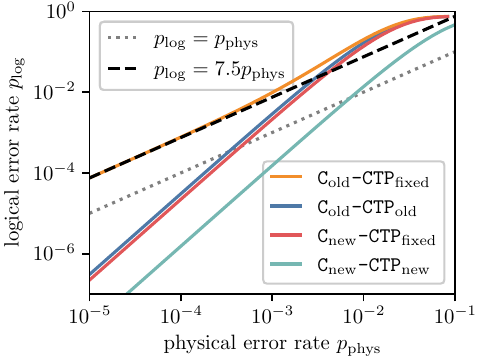}
  \caption{
    Probability of a logical failure per QEC code cycle ($p_{\mathrm{log}}$) as a function of the physical error rate ($p_{\mathrm{phys}}$) for different combinations of the circuits in Figure \ref{fig:circuits_DBR} (respectively, \CO{} and \CN{}) and \CTP{}-family noise models.
    The physical error rate at which a solid line intersects the dotted line is the pseudothreshold for the corresponding circuit and noise model.
    Solid lines in the legend are sorted top-to-bottom in the same order that they appear in the plot.
  }
  \label{fig:sim_old_vs_new}
\end{figure}

To illustrate the ideas in Section \ref{sec:design}, chiefly
\begin{enumerate*}[label=(\roman*)]
  \item the fragility of the old circuit in Figure \ref{fig:circuits_DBR}(a) to phase errors on the ancilla qubits, and
  \item the FT of the new circuit in Figure \ref{fig:circuits_DBR}(b) in the presence of such errors,
\end{enumerate*}
we simulate these circuits with the \CTP{} noise models defined above.
The results of these simulations are provided in Figure \ref{fig:sim_old_vs_new}, which shows the probability $p_\log$ that a single round of error correction irreversibly corrupts a logical state of the data qubits as a function of the physical error rate $p_\phys$.

The first observation to make about Figure \ref{fig:sim_old_vs_new} is that while \CO\d\CTPO{} (the ``old'' circuit and ``old'' noise model from Ref.~\cite{crow2016improved}) crosses the $p_\log=p_\phys$ line, indicating the existence of a pseudothreshold, this pseudothreshold vanishes when the noise model additionally allows for phase errors, \CO\d\CTPF{}.
In other words, the circuit in Figure ~\ref{fig:circuits_DBR}(a) is not fault-tolerant.
In fact, the logical error rate $p_\log$ for \CO\d\CTPF{} can be estimated analytically by counting the circuit errors in Figure \ref{fig:circuits_DBR}(a) that propagate to an uncorrectable error on the data qubits.
There are seven (7) ancilla qubits, for each of which there are two (2) locations at which a phase error propagates to an uncorrectable two-qubit $XX$ or $ZZ$ error on the data qubits.
Each such location is associated with a two-qubit depolarizing channel, for which 8 out of 15 possible two-qubit Pauli errors induce a phase error on the ancilla.
In total, this counting estimate predicts that
\begin{align}
  p_\log
  = 7\cdot 2 \cdot \frac{8}{15} \, p_\phys + O(p_\phys^2)
  \approx 7.5 \, p_\phys,
\end{align}
which quantitatively agrees with the results in Figure \ref{fig:sim_old_vs_new}.

The second takeaway from Figure \ref{fig:sim_old_vs_new} is that keeping the more realistic \CTPF{} noise model and switching to the new circuit in Figure \ref{fig:circuits_DBR}(b) restores FT, as shown by the curve labeled \CN\d\CTPF{}.
Indeed, the behavior of \CN\d\CTPF{} nearly coincides with that for \CO\d\CTPO{}.
In this sense, the new circuit ``saves'' the quantitative predictions in Ref.~\cite{crow2016improved}, after correcting for an oversight in the noise model.

Finally, Figure \ref{fig:sim_old_vs_new} shows that the pseudothreshold predicted by \CN\d\CTPF{} for the MFQEC Steane code is in fact too \emph{pessimistic}, as seen by comparison with \CN\d\CTPN{}.
The essential difference between the \CTPF{} and \CTPN{} noise models boils down to the interpretation of the physical error rate $p_\phys$.
In the \CTPF{} noise model, $p_\phys$ is the probability of error for \emph{each} of four Pauli channels that get inserted after a five-qubit gate in Figure \ref{fig:circuits_DBR}(b).
Setting the probability of error for each of these channels to $p_\phys$ therefore corresponds to setting the probability of error (infidelity) for each five-qubit gate to $1-(1-p_\phys)^4\approx 4p_\phys$.
Increasing the probability of error by a factor of 4 in turn decreases the pseudothreshold at which a curve with $p_\log\sim p_\phys^2$ crosses the dotted line at $p_\log=p_\phys$ by a factor of $4^2=16$.
Altogether, we find that the MFQEC variant of the Steane code in Figure \ref{fig:circuits_DBR}(b) has a pseudothreshold of about $0.7\%$ with the \CTPN{} noise model, in which $p_\phys$ is the infidelity of each gate.

\subsection{Identifying conditions for fault tolerance}
\label{sec:conditions}

The results in Section \ref{sec:CTP_sims} show that the MFQEC Steane code in Figure \ref{fig:circuits_DBR}(b) is fault-tolerant with respect to the \CTP{} noise models.
We now relax the assumptions of the \CTP{} noise models, and identify the conditions that a Pauli noise model must satisfy to preserve the FT of the circuit in Figure \ref{fig:circuits_DBR}(b).

\begin{figure}
  \centering
  \includegraphics{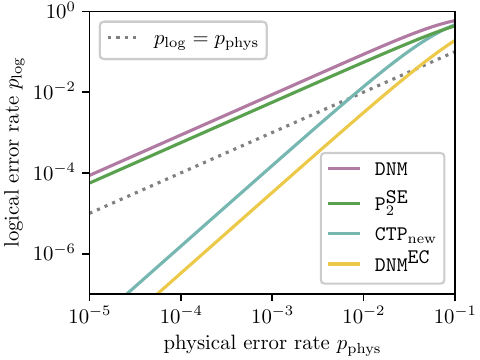}
  \caption{
    Probability of a logical failure per QEC code cycle ($p_{\mathrm{log}}$) as a function of the physical error rate ($p_{\mathrm{phys}}$) for the circuit in Figure \ref{fig:circuits_DBR}(b) with different noise models.
    The cyan curve for \CTPN{} in this figure is the same as that for \CN\d\CTPN{} in Figure \ref{fig:sim_old_vs_new}.
    Solid lines in the legend are sorted top-to-bottom in the same order that they appear in the plot.
  }
  \label{fig:sim_subcircuit_noise}
\end{figure}

To illustrate the impact of different errors, we apply the \PP{} noise model to only the syndrome extraction (\SE) subcircuits of Figure \ref{fig:circuits_DBR}(b), leaving the error correction (\EC) gates noiseless.
We denote this hybrid noise model by \PP{\SE}.
We likewise apply the \DNM{} noise model to only the \EC{} subcircuits and leave the \SE{} subcircuits noiseless, and denote this hybrid noise model by \DNM{\EC}.
For reference, we additionally consider applying either the \CTPN{} or \DNM{} noise model to the entire circuit.
The results of these simulations are provided in Figure \ref{fig:sim_subcircuit_noise}.

The primary takeaway from Figure \ref{fig:sim_subcircuit_noise} is the that \EC{} subcircuits of Figure \ref{fig:circuits_DBR}(b) are robust to \emph{all} gate errors, but the \SE{} subcircuits are ``fragile'': changing from the \CTPN{} noise model to \PP{\SE} breaks FT.
The important difference between these noise models is that \PP{\SE} allows for two-qubit errors on the data qubits.
The fact that \PP{\SE} spoils FT should therefore be unsurprising: it allows for precisely the same $XX$ and $ZZ$ errors that ultimately corrupted the logical state in Figure \ref{fig:circuits_DBR}(a).
Indeed, an exhaustive search through all possible Pauli gate errors in Figure \ref{fig:circuits_DBR}(b) reveals that, when decomposing Pauli strings into a product of $X$-type and $Z$-type errors (e.g.,~$Y_1 Z_2 Y_3 \sim X_1 X_3 \cdot Z_1 Z_2 Z_3$), two-qubit $XX$ and $ZZ$ errors on the data qubits in the \SE{} subcircuits are the \emph{only} gate errors that irreversibly corrupt a logical state.
This finding can be understood by the facts that:
\begin{enumerate*}[label=(\roman*)]
  \item tolerating single-qubit errors is a base requirement for FT,
  \item the possible four-qubit $XXXX$ and $ZZZZ$ data errors in Figure \ref{fig:circuits_DBR}(b) are stabilizers, and
  \item the possible three-qubit $XXX$ and $ZZZ$ data errors are equivalent to single-qubit errors modulo stabilizers.
\end{enumerate*}

If a two-qubit $XX$ or $ZZ$ data error after a \CNOTn{4} gate in Figure \ref{fig:circuits_DBR}(b) occurs with probability $p_\fault$, then the probability of a logical error after running one QEC code cycle is $p_\log\sim p_\fault$.
As long as $p_\fault\sim p_\phys^\alpha$, where $p_\phys$ is the infidelity of the \CNOTn{4} gate, the code will have $p_\log\sim p_\phys^\alpha$, which implies the existence of a pseudothreshold when $\alpha>1$.
Altogether, we find that the MFQEC Steane code in Figure \ref{fig:circuits_DBR}(b) is fault-tolerant as long as the physical implementation of a multi-qubit \CNOTn{4} gate induces two-qubit $XX$ and $ZZ$ errors on the target qubits (i.e., the qubits hit by \texttt{NOT}$_4$) with probability $\sim p_\phys^\alpha$ for some $\alpha>1$, where $p_\phys$ is the infidelity of the \CNOTn{4} gate.

The vulnerability of the MFQEC Steane code to $XX$ and $ZZ$ errors at the syndrome extraction stage can be generalized as follows.
Consider a QEC code with code distance $d$ and weight-$k$ stabilizers, and assume that syndromes are extracted using ancilla-controlled \CNOTk{} gates, as in Figure \ref{fig:circuits_DBR}(b).
If $k \ge d/2$, then in principle these multi-qubit gates address a sufficient number of qubits to potentially cause an uncorrectable error.
This vulnerability can therefore be avoided by using QEC codes with distance $d > 2k$, albeit at the cost of a reduced circuit-level code distance, $d_{\mathrm{circ}}<d$ \cite{bravyi2023highthreshold}.
In the case of the Steane code, there is an added symmetry between $X$-type and $Z$-type stabilizers, due to which a weight-$\ell$ bit-flip or phase-flip data error caused by a syndrome extraction gate is equivalent to a weight-$(k-\ell)$ error modulo stabilizers.
The highest effective weight of a syndrome extraction gate error is therefore $k/2$, which leaves the code only vulnerable to weight-$\ell$ errors with $d/2 \le \ell \le k/2$.
Any code with a similar symmetry is therefore vulnerable to multi-qubit errors at the syndrome extraction step when $d \le k$, i.e.~the code requires $d>k$ for FT.

\subsection{Fault tolerance with neutral atom gates}

Past works have proposed implementing multi-qubit gates natively with neutral atoms \cite{wu2010implementation, isenhower2011multibit, levine2019parallel, young2021asymmetric, pelegri2022highfidelity, farouk2023parallel}.
It is natural to ask whether these gates satisfy the requirements for FT identified in Section \ref{sec:conditions}.
As shown by the results for the \DNM{\EC} noise model in Figure \ref{fig:sim_subcircuit_noise}, the error correction subcircuits in Figure \ref{fig:circuits_DBR}(b) are robust to fully depolarizing gate errors.
This robustness is likely a generic feature of MFQEC codes with error correction circuits that use \CkNOT{}-like gates in a similar manner, because by construction these gates only address a number of qubits $\ell$ for which weight-$\ell$ errors are correctable by the QEC code in question.
The multi-qubit \CkNOT{}-like gates in Refs.~\cite{wu2010implementation, isenhower2011multibit, levine2019parallel, young2021asymmetric, pelegri2022highfidelity} are therefore compatible with FT for the circuit in Figure \ref{fig:circuits_DBR}(b), and generally likely to be compatible with FT for other MFQEC codes as well.

In contrast, the fragility of syndrome extraction to two-qubit data errors warrants a closer inspection of the \CNOTk{}-like gates in Refs.~\cite{young2021asymmetric, farouk2023parallel}.
To simplify our discussion, we consider a modified version of the protocol in Ref.~\cite{farouk2023parallel} that Hadamard-transforms target qubits to turn the \CNOTk{} gate into a \CZk{} gate.
At its core, this protocol relies on an asymmetry between control-target and target-target atom coupling strengths.
If all control-target couplings are equal and the target-target couplings are set to zero, then the protocol in Ref.~\cite{farouk2023parallel} can implement a \CZk{} gate with high fidelity, and without correlated target-target errors that would spoil FT.
However, if target-target couplings are $\epsilon g$, where $g$ is the control-target coupling strength and $0<\epsilon\ll1$, this protocol approximately implements the unitary $\CZk{}\cdot(\CkZ{})^\eta$ with exponent $\eta\gtrsim\epsilon$ (see simulation details in Appendix \ref{sec:gate_sim}).
The additional (\CkZ{})$^\eta$ gate is a coherent error that contains $ZZ$ components when expanded into Pauli strings.
This protocol is therefore incompatible with FT for the MFQEC circuit in Figure \ref{fig:circuits_DBR}(b).

The \CNOTk{} protocol in Ref.~\cite{young2021asymmetric} similarly relies on different control-target vs.~target-target couplings.
However, the relevant couplings in Ref.~\cite{young2021asymmetric} appear in a dressed basis that allows these couplings to be tuned independently using external control fields.
Compatibility with FT is therefore contingent on a guarantee that target-target couplings are zero to first order in any relevant sources of control error.
Such a guarantee seems unlikely, due to which we conclude that the protocol in Ref.~\cite{young2021asymmetric} is likely also incompatible with FT for the MFQEC circuit in Figure \ref{fig:circuits_DBR}(b).

As a final comment, we note that multi-qubit errors are equally a problem for FT in measurement-based QEC with multi-qubit gates.
Multi-qubit \CNOTk{} gates can be used in a manner similar to that in Figure \ref{fig:circuits_DBR}(b) for the syndrome extraction step of a measurement-based QEC code cycle.
However, the use of a multi-qubit gate for syndrome extraction precludes the use of flag qubit schemes \cite{chao2018quantum, chamberland2018flag, chao2020flag} for diagnosing syndrome extraction errors.
In both the measurement-based and measurement-free case, therefore, one must take care to ensure that multi-qubit gates do not induce uncorrectable data-qubit errors.
In the case of the Steane code, two-qubit $XX$ or $ZZ$ errors on the data qubits are uncorrectable, which constrains the \CNOTk{} gate protocols that are compatible with FT.
However, this constraint does not preclude the compatibility of the same gate protocols with FT for a higher-distance code as discussed at the end of Section \ref{sec:conditions}.
The takeaway message is that multi-qubit gates are not guaranteed to be a boon for QEC: one should also make sure that these gates have error characteristics that are compatible with FT for a given QEC code.

\subsection{Decomposing multi-qubit gates}

\begin{figure}
  \centering
  \includegraphics{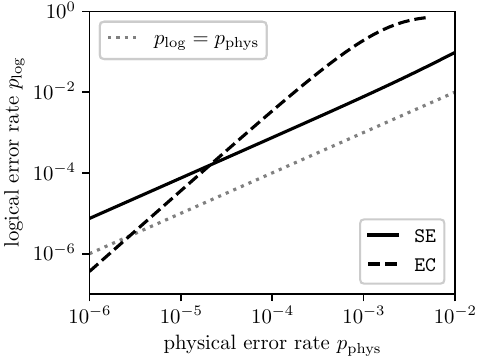}
  \caption{
    Probability of a logical failure per QEC code cycle ($p_{\mathrm{log}}$) as a function of the physical error rate ($p_{\mathrm{phys}}$) for the circuit in Figure \ref{fig:circuits_DBR}(b), with either the syndrome extraction (\SE{}) or error correction (\EC{}) subcircuits decomposed down to a two-qubit gate set.
    In each case, the decomposed subcircuits are subject to depolarizing gate errors (\DNM{}), while the non-decomposed subcircuits are noiseless.
    Due to the decomposition of multi-controlled gates into a two-qubit gate set, the decomposed \EC{} simulations in this figure do not satisfy the assumptions of the circuit simulation method presented in Section \ref{sec:simulation}.
    Logical failure rates for these circuits were therefore determined by using more computationally intensive tensor network methods to compute a transition matrix for a circuit with fixed mid-circuit errors (see Appendix \ref{sec:sampling}).
  }
  \label{fig:sim_decomps}
\end{figure}

Section \ref{sec:conditions} discusses the conditions for FT with multi-qubit gates, but it leaves open the possibility of achieving FT by decomposing multi-qubit gates down to a two-qubit gate set.
In similar spirit to the analysis in Section \ref{sec:conditions} and Figure \ref{fig:sim_subcircuit_noise}, in this section we consider decomposing either the syndrome extraction (\SE) or error correction (\EC) subcircuits of Figure \ref{fig:circuits_DBR}(b).
We decompose single-control, multi-target \CNOTk{} gates into $k$ \texttt{CNOT} gates, and we decompose multi-control, single-target \CkNOT{} and \CkZ{} gates using the ancilla-free decompositions in Ref.~\cite{hou2017quantum}.
We then apply a depolarizing (\DNM{}) noise model to the decomposed subcircuits, leaving the other subcircuits noise-free.
Note that the \CTPF{}, \CTPN{}, \PP{}, and \DNM{} noise models are equal after decomposing down to a two-qubit gate set.
The results of these simulations are provided in Figure \ref{fig:sim_decomps}.

As far as the existence of a pseudothreshold is concerned, the lesson from Figure \ref{fig:sim_decomps} is similar to that from noise model comparisons in Figure \ref{fig:sim_subcircuit_noise}, namely that the syndrome extraction (\SE{}) subcircuits are ``fragile'', while the error correction (\EC{}) subcircuits are ``robust''.
Decomposing \EC{} subcircuits preserves FT, albeit at the cost of a significantly reduced pseudothreshold.
This reduction can be accounted for by the large increase in \EC{} gate count when decomposing \CnNOT{4} gates: from 14 to 2034.
Increasing the opportunities for a single error by a factor of $\sim10^2$ should push down the pseudothreshold by a factor of $\sim10^4$.
This estimate is consistent with the reduction of the pseudothreshold for \DNM{\EC} in Figure \ref{fig:sim_subcircuit_noise} from $\sim4\times10^{-2}$ to $\sim3\times10^{-6}$ in Figure \ref{fig:sim_decomps}, as the only difference between the corresponding simulations is the decomposition of the \EC{} subcircuits.

\begin{figure}
  \centering
  \includegraphics{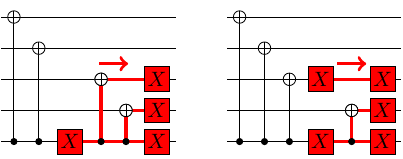}
  \caption{
    Two errors in a decomposed \CNOTn{4} gate that propagate to uncorrectable errors on the data qubits in Figure \ref{fig:circuits_DBR}(b).
    These errors are responsible for spoiling FT in the \SE{} simulations of Figure \ref{fig:sim_decomps}.
  }
  \label{fig:decomp_errors}
\end{figure}

Decomposing the \SE{} subcircuit, meanwhile, breaks FT entirely.
FT for the decomposed circuit is spoiled by essentially the same mechanism as in Figure \ref{fig:circuits_DBR}(a): it opens a location in the circuit at which a single-qubit ancilla error propagates to a two-qubit data error, shown in Figure \ref{fig:decomp_errors}.
This decomposition also allows for two-qubit ancilla-data errors that propagate to two-qubit data errors.
Analogous faults exist in Figure \ref{fig:circuits_DBR}(a) \emph{without} decomposition, but are excluded by the \CTPO{} noise model.
Altogether, decomposing multi-qubit gates down to a two-qubit gate set spoils the FT of the circuit in Figure \ref{fig:circuits_DBR}(b).

\subsection{Fault-tolerant MFQEC}

The errors spoiling FT in Figures \ref{fig:sim_old_vs_new}, \ref{fig:sim_subcircuit_noise}, and \ref{fig:sim_decomps} are generic to syndrome extraction circuits, and in fact occur in measurement-based codes as well.
In the measurement-based setting, these errors can be tolerated by the introduction of flag qubits \cite{chao2018quantum, chamberland2018flag, chao2020flag} or related schemes \cite{reichardt2020faulttolerant} to diagnose syndrome errors.
However, these techniques require repeated rounds of syndrome extraction and measurement prior to error correction.
In contrast, the measurement-free codes must extract syndromes only once before correcting errors.

One way to meet the requirements of MFQEC is through a scheme known in the measurement-based context as ``single-shot FT''.
An $r$-round extract-and-measure protocol can be converted into a single-shot protocol simply by using $r$ times the ancilla qubits, essentially performing all $r$ extraction rounds at once prior to measurement and correction.
In addition to introducing a large qubit overhead, however, in the MFQEC context this strategy requires more complex multi-qubit logic for decoding, e.g., turning \CkNOT{} gates in the error correction subcircuits into \CnNOT{\ell} gates with $\ell=k\times r$.
As an alternative, single-shot fault-tolerant QEC is known to be possible with, for example, the 4D toric code \cite{dennis2002topological} and 3D gauge color codes \cite{bombin2015singleshot}, which thereby provide a more promising avenue for MFQEC.
It may also be possible to engineer syndrome extraction schemes to circumvent the issues that we identified, for example with syndromes that span multiple rounds of error correction, as in Ref.~\cite{ercan2018measurementfree}.
We note that the discussion concerning fault-tolerant syndrome extraction in the last paragraph of Section \ref{sec:conditions} also applies to the case of two-qubit gates.

\begin{figure}
  \centering
  \includegraphics{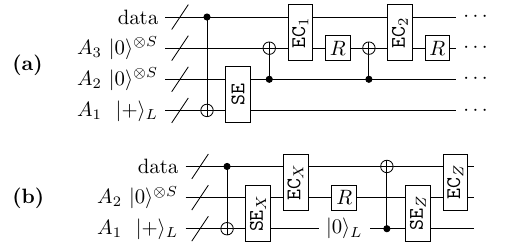}
  \caption{
    \textbf{(a)}
    Proposal in Ref.~\cite{heussen2023measurementfree} to fault-tolerantly correct bit-flip ($X$) errors in MFQEC code.
    Here $S$ is the number of syndrome bits, and $\ket{+}_L=\ket{0}_L+\ket{1}_L$ is a logical state of the QEC code.
    A transversal CNOT copies bit-flip errors onto ancilla register $A_1$, from which syndromes are extracted (\texttt{SE}) onto register $A_2$.
    These syndromes are then copied with a transversal CNOT from register $A_2$ onto register $A_3$ before applying a bit-flip correction (\texttt{EC}$_{j=1}$) to the data qubits with a multi-qubit \CkNOT{}-like gate.
    Register $A_3$ is then reset ($R$) and reused similarly for all remaining correctable bit-flip errors (indexed by $j=2,3,\cdots$).
    In practice, register $A_1$ is not needed after syndrome extraction, so its qubits can be subsequently reset and used for register $A_3$, i.e., $A_3\subseteq A_1$.
    A similar circuits corrects phase-flip ($Z$) errors.
    \textbf{(b)}
    Fault-tolerant copy-assisted MFQEC code cycle with design-based redundancy.
    Here \texttt{SE}$_P$ and \texttt{EC}$_P$ are the syndrome extraction and error correction subcircuits for correcting $P$-type errors in Figure \ref{fig:circuits_DBR}(b).
    This circuit does not need repeated reset-copy-correct cycles due to its resilience to syndrome errors.
  }
  \label{fig:circuits_FT_sketch}
\end{figure}

More recently, Ref.~\cite{heussen2023measurementfree} introduced an alternative proposal to achieve FT through the use of additional ancilla registers, as sketched in Figure \ref{fig:circuits_FT_sketch}(a).
We refer to this as the \texttt{COPY} strategy to achieve fault-tolerant MFQEC.
The pattern of interaction in a \texttt{COPY} circuit is carefully designed to limit the propagation of correlated errors to the data qubits.
Specifically, in this work we identified syndrome extraction as the ``fragile'' step at which data qubits can acquire correlated, uncorrectable errors.
Fundamentally, this vulnerability comes from the fact that each ancilla qubit interacts with many data qubits for syndrome extraction in Figure \ref{fig:circuits_DBR}.
A single fault associated with one ancilla can thereby affect multiple data qubits.
The scheme in Ref.~\cite{heussen2023measurementfree} avoids this problem by making each data qubit interact with only a single ancilla qubit prior to syndrome extraction, through a transversal CNOT.
After syndrome extraction in Figure \ref{fig:circuits_FT_sketch}(a), resetting register $A_3$ and copying the syndrome from $A_2$ before applying each correction ensures that a fault from one correction gate (\texttt{EC}$_j$) does not corrupt the syndrome stored in $A_2$, which would propagate forward to corrupt subsequent correction gates (\texttt{EC}$_{k>j}$).

\begin{figure}
  \centering
  \includegraphics{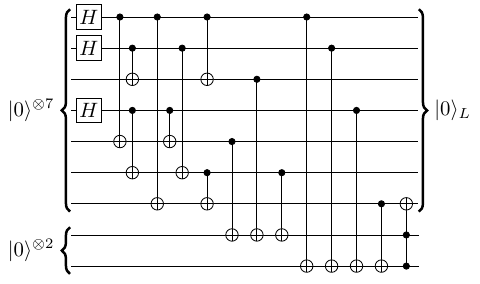}
  \caption{
    Circuit constructed in Ref.~\cite{heussen2023measurementfree} to fault-tolerantly prepare a logical $\ket{0}_L$ state of the Steane code using two ancilla qubits.
    A logical $\ket{+}_L$ state is obtained from $\ket{0}_L$ by applying a transversal Hadamard gate: $\ket{+}_L = H^{\otimes7} \ket{0}_L$.
  }
  \label{fig:circuit_0_FT}
\end{figure}

\begin{figure*}
  \centering
  \includegraphics[width=\textwidth]{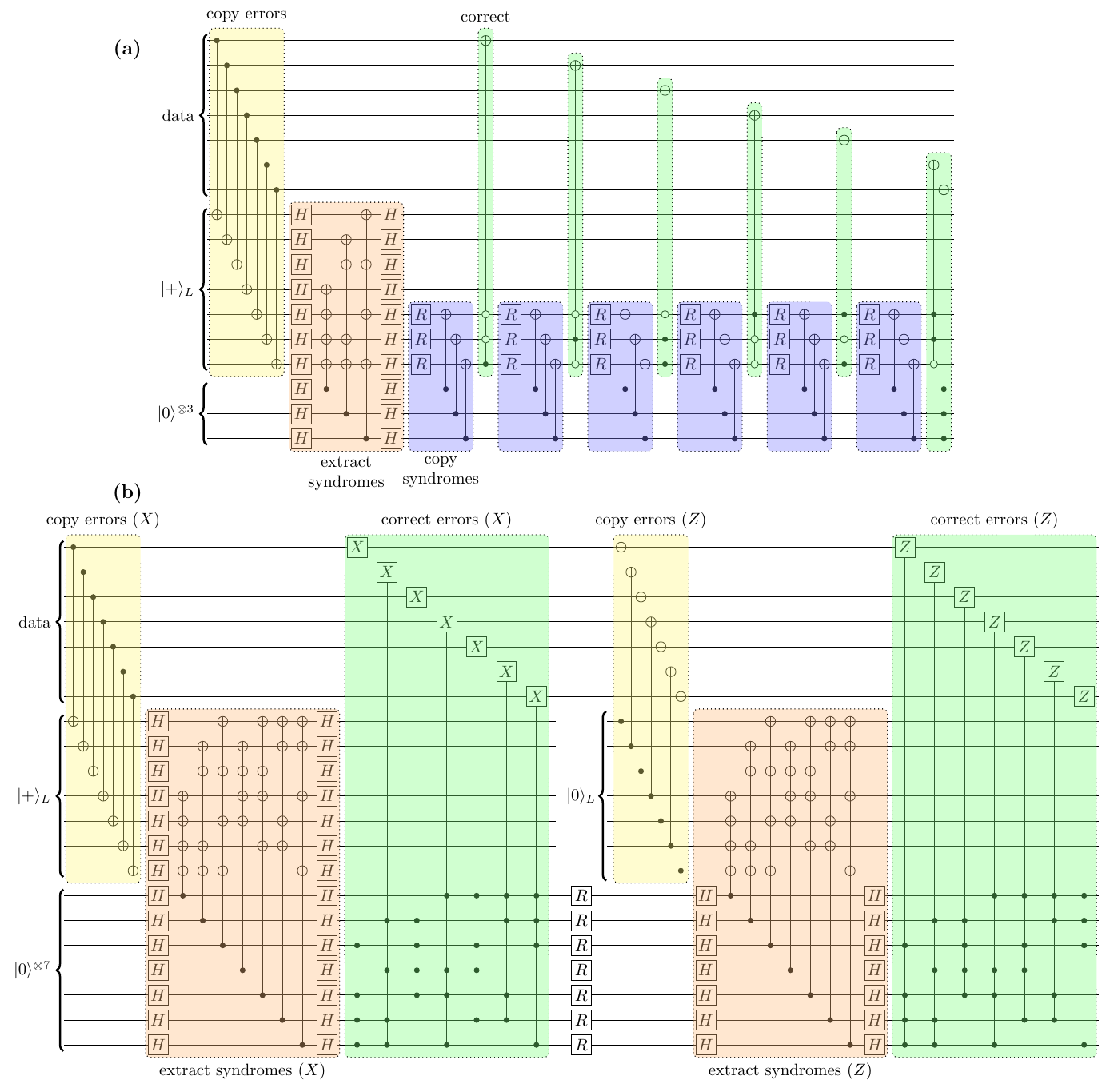}
  \caption{
    \textbf{(a)} Full circuit from Ref.~\cite{heussen2023measurementfree} for copy-assisted correction of bit-flip ($X$-type) errors in the Steane code.
    This circuit requires repeated reset-copy-correct cycles to prevent multi-qubit errors at one correction gate from spoiling the error syndrome and propagating forward to subsequent correction gates.
    \textbf{(b)} Full MFQEC code cycle achieving FT with copy-assisted syndrome extraction and design-based redundancy.
    Redundant syndrome extraction makes this circuit robust to syndrome errors, avoiding the need for repeated reset-copy-correct cycles.
    Ref.~\cite{heussen2023measurementfree} provides circuits to fault-tolerantly prepare $\ket{0}_L$ and $\ket{+}_L$.
  }
  \label{fig:circuits_FT}
\end{figure*}

In fact, the FT strategy in Ref.~\cite{heussen2023measurementfree} can be combined with design-based redundancy \cite{premakumar20212designs} (i.e., carefully chosen redundant syndrome extraction, \texttt{DBR}) to eliminate the need for multiple correction cycles in Figure \ref{fig:circuits_FT_sketch}(a), thereby reducing the total time spent on potentially slow \texttt{Reset} operations.
Figure \ref{fig:circuits_FT_sketch}(b) sketches out a complete QEC code cycle of the combined \texttt{DBR+COPY} strategy for FT.
Both circuits sketched in Figure \ref{fig:circuits_FT_sketch} rely on the fault-tolerant preparation of logical $\ket{0}_L$ and $\ket{+}_L$ states.
A circuit to fault-tolerantly construct $\ket{0}_L$ for the Steane code using two ancillas is provided in Figure \ref{fig:circuit_0_FT}, and $\ket{+}_L$ is obtained from $\ket{0}_L$ by applying a Hadamard gate to each data qubit.
Figure \ref{fig:circuits_FT} expands the circuits sketched in Figure \ref{fig:circuits_FT_sketch} into explicit gate sequences for the Steane code.
We note that Ref.~\cite{heussen2023measurementfree} also provides a circuit to correct bit-flip ($X$) errors in the Steane code with only a single round of error correction and \texttt{Reset} gates.
However, the single-round circuit in Ref.~\cite{heussen2023measurementfree} has greater qubit overhead, requiring 35 qubits total, as opposed to 21 qubits with \texttt{DBR+COPY}.

\begin{figure}
  \centering
  \includegraphics{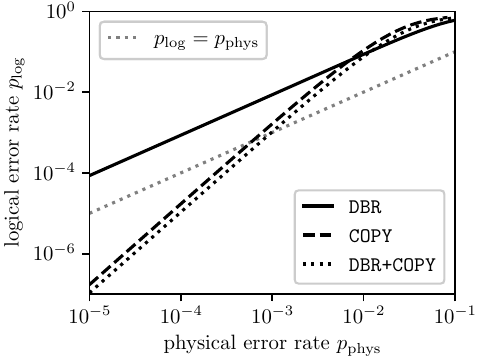}
  \caption{
    Probability of a logical failure per QEC code cycle ($p_{\mathrm{log}}$) as a function of the physical error rate ($p_{\mathrm{phys}}$) for the \texttt{DBR}, \texttt{COPY}, and \texttt{DBR+COPY} variants of the Steane code with a fully depolarizing (\texttt{DNM}) noise model.
    The \texttt{DBR} curve in this figure is the same as that for \texttt{DNM} in Figure \ref{fig:sim_subcircuit_noise}.
  }
  \label{fig:sim_copy_DBR}
\end{figure}

Altogether, the probability of logical failure per code cycle as a function of the physical error rate for the \texttt{DBR}, \texttt{COPY}, and \texttt{DBR+COPY} strategies with a depolarizing (\texttt{DNM}) noise model is provided in Figure \ref{fig:sim_copy_DBR}.
These simulations use the circuit in Figure \ref{fig:circuit_0_FT} to construct logical $\ket{0}_L$ and $\ket{+}_L$ states as necessary, resetting the ancillas for re-use later in the circuit.
In total, we find that the \texttt{COPY} scheme (which uses 17 qubits and 14 \CnNOT{3}-like gates) has a pseudothreshold of $\sim0.06\%$%
\footnote{
  This pseudothreshold is higher than that reported for the \texttt{COPY} strategy in Ref.~\cite{heussen2023measurementfree} because our use of multi-qubit gates reduces the overall number of gates in a code cycle.
}, while the \texttt{DBR+COPY} scheme (which uses 21 qubits and 14 \CnNOT{4}-like gates) has a pseudothreshold of $\sim0.1\%$.

\section{Summary and conclusions}
\label{sec:summary}

We have revisited the question of FT in a MFQEC variant of the Steane code, finding that previous work overlooked physically relevant errors that spoil FT, but that FT can be recovered by amending the strategy for syndrome extraction.
In order to investigate FT systematically, we introduced an efficient method for classically simulating MFQEC circuits, which we hope will find broader use in the study of MFQEC codes.
Using our simulation method, we found that the amended MFQEC Steane code has a pseudothreshold of $\sim0.7\%$ with a previously considered noise model that allows for correlated control-target qubit errors.

We then relaxed the assumptions of the noise model and identified general conditions for FT with multi-qubit gates.
We discovered that there are no particularly stringent requirements for the multi-control, single-target \CkNOT{}-like gates that are used for error correction, aside from achieving high gate fidelities.
However, the single-control, multi-target \CNOTk{}-like gates that are used for syndrome extraction must, in the case of the Steane code, have a vanishing probability at first order in the gate infidelity of inducing target-target $XX$ and $ZZ$ errors.
We generalized these findings to other codes as well.
Notably, we found that existing proposals for natively implementing \CNOTk{}-like gates with neutral atoms do not satisfy FT requirements for straightforward implementations of the MFQEC Steane code.
This problem applies equally to the measurement-based setting if \CNOTk{}-like gates are used for syndrome extraction.
We further found that decomposing \CNOTk{}-like gates down to a two-qubit gate set similarly spoils FT.
Finally, we discussed general requirements for fault-tolerant MFQEC with both multi-qubit and two-qubit gate sets, highlighting the role of single-shot FT, as well as a recent proposal to achieve FT by copying errors onto an ancilla register.
By combining design-based redundancy, copy-assisted FT, and multi-qubit gates, we constructed a measurement-free variant of the Steane code with a pseudothreshold of $\sim0.1\%$.

Our study highlights the fact that incorporating multi-qubit gates into QEC circuits has advantages, but does not guarantee improved code performance.
It is crucial to thoroughly examine FT conditions for QEC circuits that involve multi-qubit gates.
At the same time, proposals for natively implementing multi-qubit gates should be mindful of their potential (in)compatibility with the requirements for fault-tolerant QEC.
We hope that these lessons carry through to future studies of MFQEC and native hardware implementations of multi-qubit gates.

\begin{acknowledgments}
  We thank D.~Crow for helpful discussions.
  This research was sponsored in part by the Army Research Office (ARO) under Grant Number W911NF-17-1-0274.
  The views and conclusions contained in this document are those of the authors and should not be interpreted as representing the official policies, either expressed or implied, of the Army Research Office (ARO), or the US Government.
  The US Government is authorized to reproduce and distribute reprints for Government purposes notwithstanding any copyright notation herein.
  MS was supported by NSF PHY-1720220.
\end{acknowledgments}

\bibliography{main}

\appendix

\section{Sampling algorithm to compute logical error rates}
\label{sec:sampling}

Section \ref{sec:simulation} provides a method to propagate Pauli errors through a MFQEC circuit to determine an overall Pauli error on the data qubits at the end of a MFQEC code cycle.
Here, we describe a sampling algorithm that averages over cirucit errors to compute logical error rates $p_\log$ for a continuous range of physical error rates $p\le p_{\mathrm{max}}$.
The algorithm presented here was used for the FT analysis in Section \ref{sec:fault_tolerance}.
We note that the key technique in our algorithm -- importance sampling \cite{tokdar2010importance} -- is by no means novel, and is indeed standard practice in the QEC literature \cite{li2017fault, trout2018simulating}.
Even so, we describe our algorithm below for completion, with adaptations for our present work.

At a high level, our strategy will be to write the logical error rate $p_\log$ as a function of a \emph{transition matrix} $T$ that captures the essential error correction properties of a MFQEC circuit.
If we assume a stochastic noise model in which different circuit errors occur independently with probability $p$, we can expand the transition matrix into a series of the form $T = \sum_{w\ge0} \beta(w, p) \bar{T}(w)$, where the scalar coefficients $\beta(w, p)\sim p^w$, and the matrices $\bar{T}(w)$ can can be estimated by simulating MFQEC circuits with $w$ randomly chosen circuit errors.
Crucially, the matrices $\bar{T}(w)$ are independent of $p$, and the series expansion for $T$ converges rapidly when $p\ll1$, which means that estimating $\bar{T}(w)$ for a few values of $w$ allows us to estimate the transition matrix $T$, and in turn the logical error rate $p_\log$, for a continuous range of physical error rates $p$.

We begin by identifying, for a given QEC code that encodes a single logical qubit into the state of many data qubits, the group $E$ of all data-qubit Pauli strings modulo stabilizers and global phase.
That is, two Pauli strings $P_1$ and $P_2$ are identified in $E$ iff $P_1 = e^{\ii\theta} P_2 S$ for some real angle $\theta$ and stabilizer $S$, in which case applying either of $P_1$ or $P_2$ to a logical state $\ket\psi$ results in the same state up to global phase: $P_1\ket\psi = e^{\ii\theta} P_2 S \ket\psi = e^{\ii\theta} P_2 \ket\psi$.
For a given MFQEC circuit and fixed noise model, we denote the probability that the circuit converts the error $e\in E$ on the data qubits into the error $f\in E$ by $T_{fe}$, and collect these probabilities into the transition matrix $T = \sum_{e,f \in E} T_{fe} \op{f}{e}$.
We then define the logical error rate of a given MFQEC circuit and noise model as the probability with which a single code cycle irreversibly corrupts a logical state of the data qubits, which can be written as
\begin{align}
  p_\log = 1 - \sum_{f\in E_\cor} T_{f e_0},
\end{align}
where $e_0$ is the ``trivial'' error corresponding to an all-identity Pauli string, and $E_\cor\subset E$ is the set of correctable errors for the code.
In the case of the Steane code, $E_\cor$ consists of the trivial error $e_0$ together with the set of single-qubit Pauli errors (modulo stabilizers and global phase).

Given a circuit with $n$ error sites, we can specify a fixed set of circuit errors by an integer array $k=(k_1,k_2,\cdots,k_n)$ in which the value of $k_j$ uniquely identifies the error at site $j$, with $k_j=0$ indicating no error.
Note that a single ``site'' $j$ may be identified with several qubits, such that some values of $k_j$ may correspond to multi-qubit errors.
Letting $\gamma(k)$ denote the probability of occurrence for the errors specified by $k$, we can expand
\begin{align}
  T = \sum_k \gamma(k) T(k),
  \label{eq:trans_error}
\end{align}
where $T(k)$ is the transition matrix for the circuit with a fixed choice of errors, $k$.
For a circuit and noise model amenable to the simulation method in Section \ref{sec:simulation}, each error $e\in E$ propagates through the circuit with mid-circuit errors $k$ to another error $\tilde{e}(k)\in E$, so the matrix entries of $T(k)$ are $T_{fe}(k) = \delta_{f,\tilde{e}(k)}$, where $\delta$ is the Kronecker delta.

The transition matrix $T$ can be estimated by sampling errors $k$ according to the probability distribution $\gamma(k)$, and averaging over the corresponding matrices $T(k)$.
However, the distribution $\gamma(k)$ depends on the probability $p$ of individual circuit errors, so this procedure requires re-sampling for each value of $p$.
The remainder of this section is devoted to obtaining a sampling algorithm for computing $T$ that ``recycles'' samples of $k$ in such a way as to allow computing $T$ for any physical error rate $p\le p_{\mathrm{max}}$.

To this end, we let $\gamma_j(\ell)$ denote the probability of error $\ell$ on site $j$, and assume that errors occur independently at each site with probability $p$, such that $\sum_\ell\gamma_j(\ell) = 1$ and $\gamma_j(0)=1-p$.
The probability of occurrence for the errors specified by $k$ is then
\begin{align}
  \gamma(k)
  = \prod_j \gamma_j(k_j)
  = p^{\abs{k}} (1-p)^{n-\abs{k}} q(k),
  \label{eq:prob_error}
\end{align}
where $\abs{k}$ is the number of nonzero entries in $k$, and
\begin{align}
  q(k) = \prod_{j: k_j\ne 0} q_j(k_j),
  &&
  q_j(\ell) =
  \begin{cases}
    \gamma_j(\ell) / p & \ell \ne 0 \\
    0                  & \ell = 0
  \end{cases}.
\end{align}
Here $q_j(\ell)$ is the probability that if an error occurs on site $j$, that error is $\ell$.
The expansion of $\gamma(k)$ in Eq.~\eqref{eq:prob_error} motivates collecting together terms in Eq.~\eqref{eq:trans_error} with $w=\abs{k}$ errors, arriving at the expansion
\begin{align}
  T = \sum_{w=0}^n \beta(w, p) \bar{T}(w),
  \label{eq:trans_weight}
\end{align}
where $\beta(w, p) = p^w (1-p)^{n-w} {n \choose w}$ is a binomial distribution, and
\begin{align}
  \bar{T}(w)
  &= {n \choose w}^{-1} \sum_{k:\abs{k}=w} q(k) T(k) \\
  &= \EE_{J:\abs{J}=w} \EE_{k:k\sim J}^q T(k).
  \label{eq:trans_weight_sampling}
\end{align}
Here $\EE_{J:\abs{J}=w}$ denotes a (uniform) average over the ${n\choose w}$ possible choices of $w$ lattice sites, i.e.~choices of $J\subset\set{1,2,\cdots,n}$ with $\abs{J}=w$, and $\EE_{k:k\sim J}^q$ denotes an $q$-weighted average over choices of nontrivial errors on the sites in $J$, i.e.~over choices of $k$ with $k_j\ne0$ iff $j\in J$ (denoted by $k\sim J$).
If nontrivial errors are chosen uniformly at each error site, as e.g.~in the case of a depolarizing noise model (and all noise models considered in Section \ref{sec:fault_tolerance}), then $q(k)$ is a uniform probability distribution over $k\sim J$.

The expression of $\bar{T}(w)$ as a weighted average in Eq.~\eqref{eq:trans_weight_sampling} motivates the following algorithm for estimating matrix elements of $\bar{T}(w)$:
\begin{enumerate}[label=\thesection.\arabic*., ref=\thesection.\arabic*]
  \item \label{step:pick_sites}
    Pick a uniformly random choice of $w$ error sites, $J$.
  \item For each site $j\in J$ pick an error $k_j$ according to the probability distribution $q_j$.  These choices define a collection of circuit errors, $k$.
  \item \label{step:simulate_circuit}
    For each initial error $e\in E$ of interest (clarified below), use the algorithm in Section \ref{sec:simulation} to propagate the error $e$ through a circuit with errors $k$, thereby obtaining a new error $\tilde{e}(k)\in E$.
    Set $T_{fe}(k)=\delta_{f,\tilde{e}(k)}$.
  \item Repeat steps \ref{step:pick_sites}--\ref{step:simulate_circuit} a total of $N_w$ times, and take the average over computed values of $T_{fe}(k)$, thereby obtaining an estimate of $\bar{T}_{fe}(w)$.
\end{enumerate}
For the purposes of computing the logical error rate $p_\log$, the only initial error of interest at Step \ref{step:simulate_circuit} is the trivial error $e_0$.
The only question remaining is that of choosing $N_w$, the number of times that one should sample circuit errors to estimate $\bar{T}(w)$.

For a fixed physical error rate $p$, the central limit theorem guarantees that drawing $N$ circuit errors $k$ according to the probability distribution $\gamma(k)$ and averaging the corresponding values of $T(k)$ provides an estimate of $T$ with relative error $\sim1/\sqrt{N}$.
Eqs.~\eqref{eq:trans_weight} and \eqref{eq:trans_weight_sampling} tell us that we can sample $k$ with probability $\gamma(k)$ by following steps \ref{step:pick_sites}--\ref{step:simulate_circuit} above.
In expectation, this procedure would essentially use $N\times \beta(w, p)$ samples to estimate $\bar{T}(w)$.
However, the distribution $\beta(w, p)\sim p^w$ vanishes as $p\to0$, which means that in practice this procedure would spend most of its time ``sampling'' the unique \mbox{(non-)error} $k_0=(0, 0, \cdots, 0)$ with $w=0$.
A better idea is therefore to compute $\bar{T}(0) = T(k_0)$ once, and devote the $N$ samples to estimating matrices with $w>1$, which translates to a relative error of $\sim1/\sqrt{N}$ in the estimated value of $p_\log$.
In order to preserve the guarantee on relative error for all physical error rates $p\le p_{\mathrm{max}}$, we therefore set
\begin{align}
  N_w = \round\left(N \times
  \max_{p\le p_{\mathrm{max}}} \frac{\beta(w, p)}{p_{\mathrm{error}}}
  \right),
\end{align}
where $\round(x)$ rounds $x$ to the nearest integer, and $p_{\mathrm{error}} = \sum_{w>0} \beta(w, p) = 1 - \beta(0, p)$ is the probability that at least one circuit error occurs.
We set $N = 10^4$ for the simulations in Section \ref{sec:fault_tolerance}.

\section{Neutral atom gate simulations}
\label{sec:gate_sim}

Here we describe our simulations of the \CNOTk{} gate protocol in Ref.~\cite{farouk2023parallel}.
This protocol involves $k+1$ atoms with four internal states: $\ket{0}$, $\ket{1}$, $\ket{r}$, and $\ket{p}$.
Here $\ket{0}$ and $\ket{1}$ are the computational (qubit) states, $\ket{r}$ is a highly-excited Rydberg state, and $\ket{p}$ is an auxiliary intermediate state.
To simplify our analysis, we consider a modified version of the protocol in Ref.~\cite{farouk2023parallel}, in which the target qubits of the \CNOTk{} gate are Hadamard-transformed, altogether implementing a \CZk{} gate.
We also neglect all sources of error, such as control error or decay from the Rydberg state.

\subsection{The protocol}

The revised protocol that we simulate is as follows:
\begin{enumerate}[label=\thesection.\arabic*., ref=\thesection.\arabic*]
  \item \label{step:pi_1r}
    Apply a $\pi$-pulse to swap the $\ket{1}$ and $\ket{r}$ states of the control atom, leaving the target atoms unaffected.
  \item \label{step:pulse}
    Evolve for time $T$ under the time-dependent Hamiltonian
    \begin{align}
      H(t) = H_{\mathrm{int}} + \sum_{j\in\text{targets}} H_{\mathrm{pulse}}^{(j)}(t),
    \end{align}
    where $H_{\mathrm{int}}$ is a two-body interaction Hamiltonian, $H_{\mathrm{pulse}}^{(j)}$ is the single-atom Hamiltonian $H_{\mathrm{pulse}}$ applied to atom $j$, and the index $j$ runs over the target atoms of the
    \CZk{} gate.
    We define $H_{\mathrm{int}}$ and $H_{\mathrm{pulse}}$ below.
  \item Apply a $\pi$-pulse to swap the $\ket{1}$ and $\ket{r}$ states of the control atom (identically to Step \ref{step:pi_1r}).
  \item \label{step:correct}
    Apply corrective \RZ{} gates (clarified below).
\end{enumerate}
The interaction Hamiltonian that appears in Step \ref{step:pulse} is
\begin{align}
  H_{\mathrm{int}} = \sum_{\text{atoms}\,i<j} g_{ij} \op{rr}_{ij},
\end{align}
where $g_{ij}$ is the Rydberg-Rydberg interaction strength between atoms $i$ and $j$.
In an ideal case, the inter-atomic coupling strength $g_{ij}=g$ iff one of $i,j$ is a control atom and the other a target atom, and $g_{ij}=0$ otherwise.
The single-atom Hamiltonian $H_{\mathrm{pulse}}$ in Step \ref{step:pulse} is
\begin{align}
  H_{\mathrm{pulse}}(t)
  = f(t) \Om{p} s_{\mathrm{x}}^{(0,p)}
  + \Om{c} s_{\mathrm{x}}^{(p,r)}
  - \Delta \op{p}{p},
\end{align}
where $f(t) = \sin(\pi t/T)^2$, the operator $s_{\mathrm{x}}^{(a,b)} = \frac12 (\op{a}{b} + \op{b}{a})$ is a spin-$x$ operator in the $\set{\ket{a},\ket{b}}$ subspace of an atom, and the detuning $\Delta = \frac{3}{32\pi} \Om{p}^2 T$%
\footnote{
Note that in addition to factoring out the time dependence of $\Om{p}$ into $f(t)$, our definition of $\Om{p}$ is larger than that in Ref.~\cite{farouk2023parallel} by a factor of $\sqrt{2}$.
That is, $\Om{p}^{\text{ours}} = \sqrt{2} \times \Om{p}^{\mathrm{theirs}}$.
}.
We provide the parameters used in our simulations in Table \ref{tab:params}.

\begin{table}
  \caption{
    \CZk{} gate protocol parameters.}
  \begin{tabular}{|c|c|}
    \hline
    $T$      & $1~\mu$s             \\ \hline
    $\Om{p}$ & $100\times 2\pi$ MHz \\ \hline
    $\Om{c}$ & $200\times 2\pi$ MHz \\ \hline
    $g$      & $500\times 2\pi$ MHz \\ \hline
  \end{tabular}
  \label{tab:params}
\end{table}

The last point that needs clarification is Step \ref{step:correct} of the protocol.
This step applies the gate
\begin{align}
  \RZ{}(\bm\theta) = \prod_j \RZ{}_j(\theta_j),
\end{align}
where
\begin{align}
  \RZ{}_j(\theta_j) = \exp(-\ii\theta_j s_{\mathrm{z}}^{(0,1)})
\end{align}
is a single-qubit phase gate defined by the spin-$z$ operator $s_{\mathrm{z}}^{(0,1)} = \frac12 (\op{0}{0}_j - \op{1}{1}_j)$ on the $\set{\ket{0}, \ket{1}}$ subspace of atom $j$, and the angles $\theta_j$ are determined by maximizing the entanglement fidelity \cite{nielsen1996entanglement, nielsen2002simple} of the protocol unitary $U(\bm\theta)$ with a target unitary $U_{\text{target}}$ in the computational subspace.
The entanglement fidelity between $U(\bm\theta)$ and $U_{\text{target}}$ can be written as
\begin{align}
  F(\bm\theta)
  = \F(U_{\text{target}}, U(\bm\theta))
  = \abs{\frac1d\Tr_{\text{comp}}[U_{\text{target}}^\dag U(\bm\theta)]}^2,
\end{align}
where $d=2^{k+1}$ is the dimension of the computational subspace, and $\Tr_{\text{comp}}$ denotes a trace over this subspace, i.e.~over the subspace of computational states $\ket{b}\in\set{\ket{0},\ket{1}}^{\otimes(k+1)}$.

\subsection{The ideal case: zero inter-target couplings}

Using ideal couplings and $U_{\text{target}} = \CZk{}$, we find infidelities $I(\bm\theta_\star) \sim 10^{-5}$ for $k\in\set{1,2,3,4}$, where $I(\bm\theta) = 1 - F(\bm\theta)$ and $\bm\theta_\star = \argmax_{\bm\theta} F(\bm\theta)$ is the collection of angles that maximizes $F(\bm\theta)$.
To determine the probability of a Pauli error $P$, we compute the entanglement fidelity $\F(P U_{\text{target}}, U(\bm\theta_\star))$ \cite{perlin2023short}.
We thereby find probabilities on the scale of $\sim10^{-7}$ for $ZZ$ errors on control-target qubit pairs, which can be tolerated by the circuit in Figure \ref{fig:circuits_DBR}(b).

The infidelity of $\sim10^{-5}$ is mostly attributed to stray (nonzero) populations of atomic Rydberg states at the end of the protocol, i.e.~leakage errors.
In principle, these leakage errors can be converted into depolarizing errors by letting Rydberg states decay and pumping back into the computational subspace.
Assuming perfect conversion of leakage errors to depolarizing errors, after applying corrective \RZ{} gates with angles $\bm\theta_\star$ we find probabilities on the scale of $\sim10^{-6}$ for various single-qubit errors, as well as two-qubit errors on control-target qubit pairs.
We also find a net probability $p_\fault^\star\sim 10^{-9}$ for the occurrence of an uncorrectable error.
Altogether, we conclude that the ideal \CZk{} gate is theoretically compatible with the requirements for FT found in Section \ref{sec:conditions}.

\subsection{The realistic case: nonzero inter-target couplings}

\begin{figure*}
  \centering
  \includegraphics{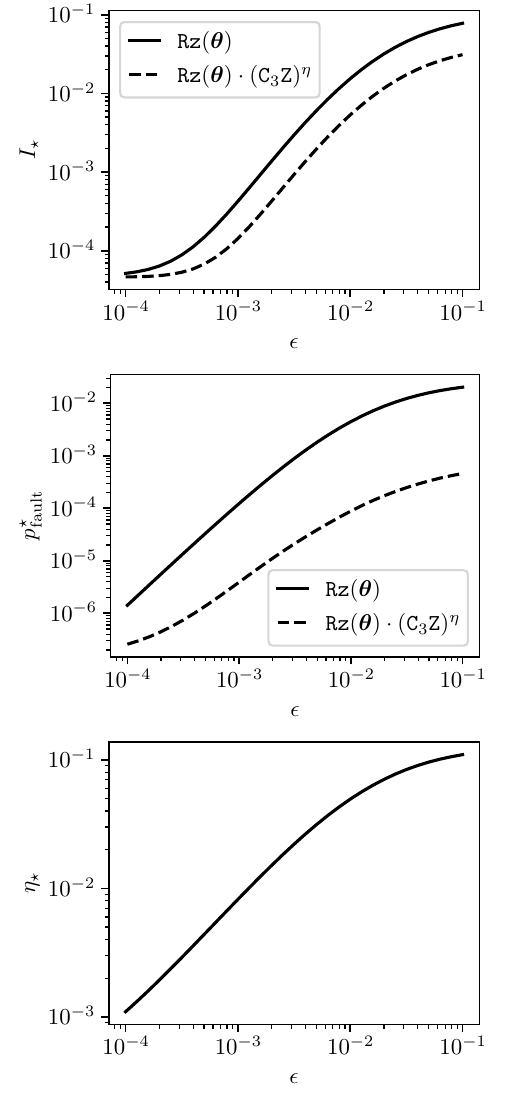}
  \caption{
    Infidelity $I_\star$, probability $p_\fault^\star$ of a fault (uncorrectable error), and optimized \CnZ{3} exponent $\eta_\star$ as a function of the target-target to control-target coupling ratio $\epsilon$ for a \CZn{3} gate protocol.
    Infidelity is minimized over corrective \RZ{} rotation angles $\bm\theta$, and simultaneously over the exponent $\eta$ of a corrective $(\CnZ{3})^\eta$ gate when applicable.
  }
  \label{fig:sim_gate_errors}
\end{figure*}

We now consider setting the target-target coupling strength to $\epsilon g$ for $0<\epsilon\ll1$.
For simplicity, we restrict our analysis to a four-qubit gate protocol, setting $k=3$.
We also convert leakage errors into depolarizing errors as discussed above, but note that this conversion will have no effect on the following results because leakage will no longer be the dominant source of error.
We then plot, for a range of coupling ratios $\epsilon$, the infidelity $I_\star$ and probability $p_\fault^\star$ of a fault (uncorrectable error) after maximizing fidelity over corrective \RZ{} rotation angles $\bm\theta$ (for each value of $\epsilon$).
In addition, we consider simultaneously optimizing over the angles $\bm\theta$ and the exponent $\eta$ in a net corrective gate $\RZ{}(\bm\theta)\cdot(\CkZ{})^\eta$.
We provide infidelities $I_\star$, probabilities of fault $p_\fault^\star$, and optimal exponents $\eta_\star$ as a function of $\epsilon$ in Figure \ref{fig:sim_gate_errors}.

Altogether, we find that the optimal \CkZ{} exponents $\eta_\star$ are large compared to the coupling ratio $\epsilon$, and that including a corrective (\CkZ{})$^\eta$ gate generally reduced the infidelity of the protocol by a factor of 3.
More strikingly, including a corrective (\CkZ{})$^\eta$ gate reduces the probability of a fault by orders of magnitude, indicating that this gate is predominantly responsible for faults.
In principle, it may be possible to construct a protocol that somehow ``echoes out'' the undesired (\CkZ{})$^\eta$ gate while keeping the \CZk{}{} gate intact.
In the absence of such a scheme, however, we conclude that the \CZk{} protocol in Ref.~\cite{farouk2023parallel} is not compatible with FT.
We do note, however, that if (\CkZ{})$^\eta$ gates are cancelled out, theoretically achievable probabilities of error ($p_\fault^\star$) may be low enough for useful QEC even if the formal conditions for FT are not satisfied.

\end{document}